\documentclass{article}
\usepackage[utf8]{inputenc}

\usepackage{indentfirst}
\usepackage{graphicx}
\usepackage{amssymb}
\usepackage{amsmath}
\usepackage[a4paper, total={6in, 8in}]{geometry}

\title{\textbf{Zero Knowledge Identification and Verification for Voting Systems}}
\author{Arunava Gantait(BS1917)\\ Rajit Goyal(BS1919) \\ Syed Sajid Husain Rizvi (BS1922) \\ Zaira Haram(BS1819)}

\begin{document}

\maketitle
\tableofcontents
\newpage

\section{Abstract}
\emph{Current methods of voter identification, especially in India, are highly primitive and error-prone, depending on verification by (mostly) sight, by highly trusted election officials. This paper attempts to provide a trustless and zero-knowledge method of voter identification, while simultaneously reducing error. It also proposes a method for vote verification, that is, ensuring that the vote cast by a legal voter is registered as cast and tallied as registered. While numerous methods of zero-knowledge identification are available in the literature, very few of those are implementable on a large scale and subject to the type of constraints that are present, eg., in India. This paper attempts to provide a solution which, while preserving the integrity of the available methods, will also be more scalable and cost-effective.}

\textbf{Keywords:} zero-knowledge, identification, verification, zk-SNARK, DRE-ip 

\section{Introduction}
The current election system in India consists of the following scheme for voter identification: at a polling booth, the polling agent has a pre-prepared sheet in which voter's low resolution photo, name and address are mentioned.\\
To verify a voter, the polling agent matches that photo with the voter's face merely with his eyes, verifies his/her name and address of the voter, takes his/her signature and allows the voter to vote. Also to ensure that a person gets to vote only once, an officer applies an ink mark on nail of the index finger for future reference.\\
The issues with this verification system are twofold: either party -  the voter or the election officer (the verifier) could be an adversary. Any adversary with sufficient technological and/or political power would be able to masquerade as another person and vote in his/her stead; conversely, the officer could have particular political leanings and coerce the voter to vote for the party of his choice (\textbf{voter coercion}). Not to mention the fact that the information at the disposal of the officer is highly sensitive and can be used for other nefarious purposes.\\
To eliminate such possibilities, we propose a method of verification using zero knowledge proof in which a voter's identity of any kind won't be revealed to anyone during the verification phase and still voter will be able to prove that he/she is an eligible voter to vote at that voting station.\\
In this report, we also propose a method in which a voter will be able to verify if their vote has been cast as intended, recorded as cast. Furthermore, anyone would be able to verify if all voters are tallied as voted.

\subsection{Zero - Knowledge Proof}
Zero-knowledge proof is a concept from cryptography in which an interactive/non-interactive method is used by one party to prove to another that a statement is true, without revealing anything other than the veracity of the statement.\\
 The essence of zero-knowledge proofs is that it is trivial to prove that one possesses knowledge of certain information by simply revealing it; the challenge is to prove such possession without revealing the information itself or any additional information.\\
 A zero-knowledge protocol should satisfy three properties:
\begin{itemize}
    \item \textbf{Completeness} - If the statement is true then a prover can convince a verifier.
    \item \textbf{Soundness} - A cheating prover can not convince a verifier of a false statement.
    \item \textbf{Zero-knowledge} - The interaction only reveals if a statement is true and nothing else.
\end{itemize}

\subsection*{Hashing}
A cryptographic hash function (CHF) is a mathematical algorithm that maps data of an arbitrary size (often called the "message") to a bit array of a fixed size (the "hash value", "hash", or "message digest"). It is a one-way function, that is, a function for which it is practically infeasible to invert or reverse the computation.\\
Ideally, the only way to find a message that produces a given hash is to attempt a brute-force search of possible inputs to see if they produce a match, or use a rainbow table of matched hashes.\\
In theoretical cryptography, the security level of a cryptographic hash function has been defined using the following properties:
\begin{itemize}
    \item \textbf{Pre-image resistance} - Given a hash value $h$, it should be difficult to find any message m such that $h = hash(m)$. This concept is related to that of a one-way function. Functions that lack this property are vulnerable to pre-image attacks.
    \item \textbf{Second pre-image resistance} - Given an input $m_1$, it should be difficult to find a different input $m_2$ such that $hash(m_1) \hspace{0.5mm} = \hspace{0.5mm} hash(m_2)$. This property is sometimes referred to as weak collision resistance. Functions that lack this property are vulnerable to second pre-image attacks. 
    \item \textbf{Collision Resistance} - It should be difficult to find two different messages $m_1$ and $m_2$ such that $hash(m_1) \hspace{0.5mm} = \hspace{0.5mm} hash(m_2)$. Such a pair is called a cryptographic hash collision.
\end{itemize}
Examples of such functions are SHA-256, SHA3-256, which transform arbitrary input to 256-bit output.
\begin{table}[h]
    \centering
    \begin{tabular}{|c|c|c|}
    \hline
    \textbf{Function} & \textbf{Input} & \textbf{Output}\\
    \hline
    SHA-256 & Hello & 185f8db32271fe25f561a6fc938b2e264306ec304eda518007d1764826381969 \\
    \hline
    SHA3-256 & Hello & 8ca66ee6b2fe4bb928a8e3cd2f508de4119c0895f22e011117e22cf9b13de7ef\\
    \hline
    \end{tabular}
    \caption{Hash Functions} 
    \label{tab:my_label}
\end{table}
\subsection{ZK-SNARK}
The acronym zk-SNARK stands for “Zero-Knowledge Succinct Non-Interactive Argument of Knowledge,” and refers to a proof construction where one can prove possession of certain information, e.g. a secret key, without revealing that information, and without any interaction between the prover and verifier.\\
“Succinct” zero-knowledge proofs can be verified within a few milliseconds, with a proof length of only a few hundred bytes even for statements about programs that are very large. In the first zero-knowledge protocols, the prover and verifier had to communicate back and forth for multiple rounds, but in “non-interactive” constructions, the proof consists of a single message sent from prover to verifier. Currently, the most efficient known way to produce zero-knowledge proofs that are non-interactive and short enough to publish to a block chain is to have an initial setup phase that generates a common reference string shared between prover and verifier.

\subsection{Merkle Tree}
A Merkle tree is a data structure in which every node, called ``leaf" is labelled with a cryptographic hash of a data block (say, a transaction) and every node that is not a leaf, called a ``branch", is labelled with the cryptographic hash of the labels of its child nodes. Also known as binary hash trees, Merkle trees allow efficient and secure verification of the contents of a large data structure. An over simplified visualization of the structure is, as follows. An average block contains over $500$ transactions, not eight.\vspace{1 cm}\\
\includegraphics[width=0.7\textwidth]{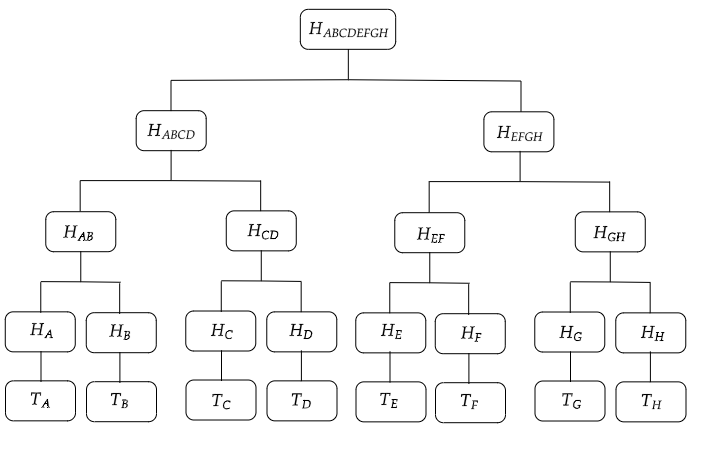}\vspace{0.5 cm}\\

\section{Commitment Phase}
Before voting, the voter would have to register himself on an app via mobile number(OTP based verification) and once registered, biometric authentication would be used for future access of the app.\\
Once the registration is done, the system will first fetch Voter ID of the person from database of eligible voters and a secret key (which will be referred to as nullifier from now on), unique to every person, will be generated and shown to person but won't be stored anywhere(neither in app nor in database). The user would need to store it somewhere privately for future use.\\
Next, hash corresponding to concatenated voter ID and nullifier will be generated which we will call \textbf{identity secret}. Again hashing the identity secret will give us another hash which we will call \textbf{identity commitment} and this will be added as a leaf of the Merkle tree.   \\
This will be done for all the voters.\\
To create merkle tree, we would  need number of leaves to be of the form $2^n, n \in \mathbb{Z}$. If we don't have enough entries, we can always always introduce hash of random entries from our side.\\

\section{Verification Phase}
\includegraphics[scale = 0.7]{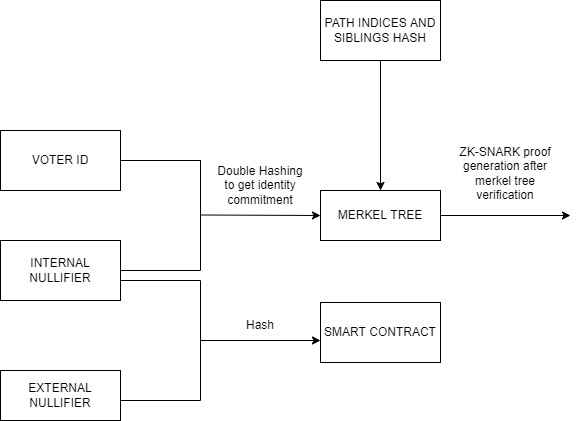}
\subsection{Public and Private Variables}
The implementation would require various inputs some of which would be coming from users, some would be picked up by the system itself and certain information would be made public which is used in the process.
    \subsection*{Public Variables}
    \begin{itemize}
        \item External Nullifiers
        \item Merkle Tree
    \end{itemize}
    \subsection*{Private Variables}
    \begin{itemize}
        \item Voter ID
        \item Internal Nullifier
        \item Path Indices and Siblings Hash
    \end{itemize}

\subsection{Merkle Tree Setup}
\indent Each identity secret (denoted by $T$) is hashed (stored as $H$), then each pair of transactions is concatenated and hashed together, and so on until there is one hash for the entire block, called the root. The root may further be combined with other information and then run through a hash function to produce a tree's unique hash. Hence, Merkle trees deal with hashes instead of the original block of transactional data.\\
\indent The Merkle tree allows users to verify a specific transaction without revealing the whole blockchain. For example, say a user wants to verify that transaction $T_D$ is included in the block (in the diagram above). Given the root hash, $H_{ABCDEFGH}$, the network is queried about $H_D$. It returns $H_C$, $H_{AB}$, and $H_{EFGH}$. Hence, everything is accounted for with three hashes: given $H_{AB}, H_C, H_{EFGH}$ and the root $H_{ABCDEFGH}$, $H_D$ (the only missing hash) has to be presented in the data. The final concatenated hash can be matched with the root to support the claim that transaction $T_D$ belongs to the block.\vspace{0.5 cm}\\
\includegraphics[width=0.7\textwidth]{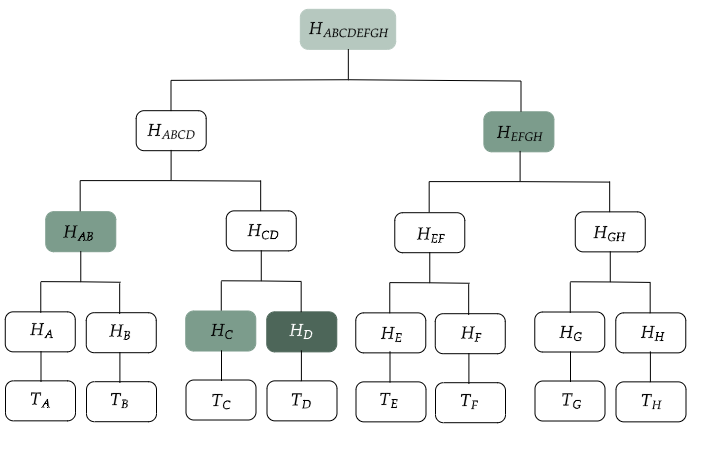}

\subsection{Verification}
\noindent\textbf{Step 1:} Voter logs into the app via fingerprint verification.\\
\textbf{Step 2:} Voter feeds in the voter ID and internal nullifier allotted to him in the commitment phase.\\
\textbf{Step 3:} The app converts the inputs into hashes, gathers information about the path indices and sibling hash and finally generates a hash which, if the voter actually has the right to vote, will match the hash at the root of the Merkle tree.\\
\textbf{Step 4:} After Merkle tree verification, if the hashes don't match, then the app will show error otherwise will generate a ZK-SNARK proof in the form a QR-code.\\
\textbf{Step 5:} At the polling station, the officer will scan the QR-code and proof will be verified which will give the polling agent a confirmation that the person is allowed to vote.\\
\textbf{Step 6:} During zero-knowledge proof verification, the system creates an hash from external nullifier (public) and the hash of the internal nullifier (stored in zero knowledge proof)  and sends it to smart contract which makes sure that it has not seen the hash before and then stores it, thus ensuring that no voter votes twice.\\

\subsection{Proofs Created by ZK-SNARK}
In this part, we will discuss what kind of proofs ZK-SNARK  produces whose probability of going wrong is negligible.\\
Consider the set of polynomials with degree atmost $d$. There are two persons, the prover and the verifier, both having a polynomial from the set. No one knows what polynomial the person has i.e. the prover and verifier know about their own polynomials only. Suppose the prover wants to prove to the verifier that he has the same polynomial the verifier has without revealing the polynomial.(One of the problem with revealing is that if one person reveals, the other person could simply lie that he has the same polynomial).\\
\indent Idea -  Two unequal polynomials with degree atmost $d$ could take same values for atmost $d$ number of points.\\
The verifier would ask the prover to compute the value of his polynomial at some integer value, The prover would reveal the value of polynomial at one point only (from where infinitely many polynomials could pass through). If the value matches with the value of polynomial that the verifier has at the same integer value, then the verifier can be sure with probability almost equivalent to 1 (exact expression is given below) that both the persons have same polynomial.\\
If the polynomials are equal, then they would equal for all the points, but if not equal, the two polynomials could be equal at atmost $d$ many points. Given that computers can with arbitarily large number of integers, two unequal polynomials attaining same values for a given integer could only happen when we choose an integer out of one of those $d$(assuming all intersection points are integer in the worst case scenario) the probability of which happening is $$\frac{d}{10^{77}}$$ which is almost equivalent to zero. \\
ZK-SNARKS produces such kind of proofs on polynomials (another example could be to check whether the polynomials of verifier and prover have a certain factor same or not) whose probability of going wrong is almost equivalent to zero.

\section{Time and Space Complexities}
In this section we present the time and space required for creating and verifying zero knowledge proofs by various existing protocols.\\
\begin{figure}[h]
    \centering
    \includegraphics[width = 10cm]{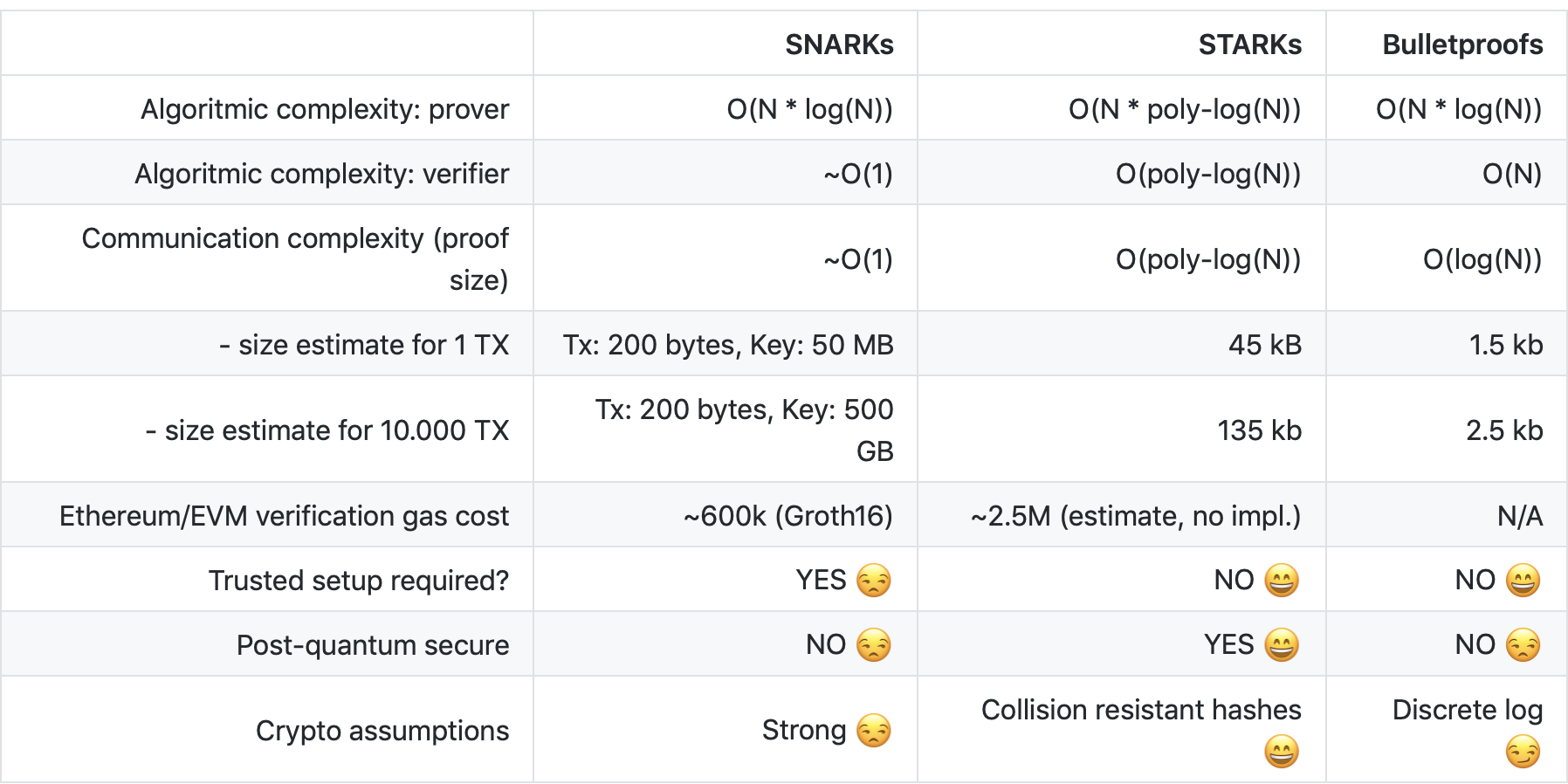}
    \label{fig:my_label}
\end{figure}

\section{Shortcomings}

While the Merkle tree protects from forgery or alteration after the creation of a record, it cannot protect from inaccurate records entered into the tree at the time of creation.

The entire protocol relies on the security of zk-SNARKS which would fail if the trusted setup for Groth16 cannot be trusted. This can be solved using multiparty computation where the randomness is supplied by multiple users participating in the Power of Tau ceremony. The setup generated by method of multiparty computation will be secure if atleast one of the party is not honest.
Recent advancements have also described methods of creating trustless setup for zk-SNARKs however that is out of the scope of this report.

\section{Alternate Implementations}
\begin{itemize}
    \item Use of zk-STARKS or Bulletproofs instead of zk-SNARKS
    \item Use of an accumulator other than merkle root such as the RSA accumulator
    \item Create a different protocol that verifies the validity of digital signatures in zero knowledge and use that to create anonymous credentials.
\end{itemize}

\section{Vote verification}
Vote verification includes assurance (to voters) that the cast votes have been (i) recorded as intended (voting phase) and (ii) tallied without influence (tallying phase). One way to work around this is make sure that if any recorded ballot is modified by an adversary before the tallying phase, it will be detected during the tallying phase. This eliminates the necessity of having a secure public storage module (bulletin board); even when the adversary has read and write access to the bulletin board and/or if voter coercion is in play, revealing the ballot receipt will be of no use to the adversary.
%This is proposed to be achieved by using Cramer-Shoup encryption to encrypt votes.
\subsection{Notations}
A non-interactive proof of knowledge of a secret $\lambda$ such that, $\Gamma=\gamma^\lambda$ for publicly known $\Gamma$ and $\gamma$, is denoted by $P_K\{\lambda:\Gamma=\gamma^\lambda\}\equiv P_K\{\lambda\}$. Also, $P_{WF}\{A:X,...,Y,Z\}\equiv P_{WF}\{A\}$ denotes the proof of well-formedness of $A$ with respect to $X,...,Y,Z$.
\subsection{Modified DRE-ip System description}
Existence of a publicly accessible bulletin board(BB) insecure enough for the adversary to have read and write access is assumed. Also, DRE is assumed to incorporate voter initiated auditing to achieve end-to-end verifiability, i.e. voters get option to audit the ballot recorded to ensure that the ballot is being prepared according to their choice. An audited ballot is not used to cast a vote. Therefore, at the end of the voting phase, the total set of ballots, $\mathbb{B}$ will be the union of the audited ballots, $\mathbb{A}$ and cast ballots, $\mathbb{C}$, i.e. $\mathbb{B}=\mathbb{A}\cup\mathbb{C}$.\\
Assuming one DRE machine used to select a candidate, the BB maintains a public blockchain per machine to store the recorded ballot sent to it. Whenever the BB receives a recorded ballot, it creates a new block and subsequently mines the block in the corresponding blockchain.
\subsection{Two candidate case}
Consider the case where there are only two candidates i.e., if $v_i$ represents the vote for the $i$-th ballot, we have $v_i\in\{0,1\}$. The algorithm for ballot encryption is as follows.
\begin{itemize}
\item[] \textit{Key Generation Phase:} In this phase, keys are generated. The key generation algorithm is executed only once prior to the voting phase.
\begin{enumerate}
\item The DRE generates an efficient description of a cyclic group $\mathbb{G}_q$ of order $q$ with two distinct random generators $g_1,g_2$.
\item It chooses five random values $(x_1,,x_2,y_1,y_2,z)$ from $\{0,1,...,q-1\}$.
\item It computes $c=g_1^{x_1}g_2^{x_2},d=g_1^{y_1}g_2^{y_2},h=g_1^z$.
\item $(c,d,h)$ is published along with the description of $G_q,q,g_1,g_2$ as its public key.
\end{enumerate}
The public key along with group descriptions are shared with the BB and published on the BB. The secret key, $(x_1,x_2,y_1,y_2,z)$ and the logarithmic relation between $g_1,g_2$ are secretly deleted by the DRE.
\item[] \textit{Initialization:} Initially, $t=0,s=0,s_1=0,m=0,n=1,n_1=1$.
\item[] \textit{Voting Phase:} Here, steps involve the DRE machine, voter and the BB.
\begin{enumerate}
    \item The voter enters the booth, initiates the voting and puts in their vote $v_i\in\{0,1\}$.
    \item A random number $r_i\in\mathbb{Z}_q^*$ is generated by the DRE. The DRE then evaluates $U_i=g_i^{r_i},V_i=g_2^{r_i},E_i=h^{r_i}g_1^{v_i},\alpha=H(U_i,V_i,E_i)$, where, $H(\cdot)$ is a universal one-way hash function.\\
    $W_i=c^{r_i}d^{r_i\alpha_i}$\\
    $P_{WF}\{E_i:g_1,g_2,c,d,h,U_i,V_i,W_i\}=P_K\{r_i:((U_i=g_1^{r_i})\land(V_i=g_2^{r_i})\land(E_i=h^{r_i})\land(W_i\newline =(cd^{\alpha_ir_i}))\lor((U_i=g_1^{r_i})\land(V_i=g_2^{r_i})\land(E_i/g_1=h^{r_i})\land(W_i=(cd^{\alpha_ir_i}))\}$\\
    $s_1=s_1+r_i,n_1=n_1U_i,P_K\{s_1:n_1=g^{s_1}\}$\\
    Here, $P_K\{r_i\}$ is a non-interactive zero knowledge proof of knowledge of $r_i$ whereas, $P_K\{s_1\}$ is a non-interactive zero knowledge proof of knowledge of sum of all random numbers generated till now i.e., $s_1$. At this stage, $s_1=\sum_{j\in\mathbb{B}}r_j,n_1=\prod_{j\in\mathbb{B}}U_j$.\\
    A signed receipt, including the unique ballot index $i$ and the ballot content $(U_i,V_i,E_i,W_i,$ $P_{WF}\{E_i\},P_K\{s_1\})$ is provided to the voter.
    \item The voter receives the first part of the receipt and decides to either audit the ballot or confirm their vote.
    \item In case of audit, the DRE adds $i$ to $\mathbb{A}$. A signed receipt of the audit, clearly marked as audited, including $r_i$ and $v_i$ is provided to the voter who keeps the receipt and verifies their choice of $v_i$. If the verification succeeds, voting continues to step 1 else, the voter should raise a dispute.\\
    The DRE then merges both parts of the receipt in a single part, $(i:(U_i,V_i,E_i,W_i,P_{WF}\{$  $E_i\},P_K\{s_1\}),(\textrm{audited},r_i,v_i)$ and creates a block to mine it in the block-chain and send the transaction to the BB.
    \item In case of confirmation, the DRE adds $i$ to $\mathbb{C}$, updates the tally, the sum and evaluates: $t=\sum_{j\in\mathbb{C}}v_j,m=\sum_{j\in\mathbb{C}}r_j\alpha_j,s=\sum_{j\in\mathbb{C}}r_j,n=\prod_{j\in\mathbb{C}}U_j$. It then evaluates $P_K\{s:n=g_1^{s}\}$, the non interactive zkp of knowledge of the partial sum $s$. A signed receipt clearly marked as confirmed and including $P_K\{s\}$ is provided to the voter. The DRE securely deletes both $r_i$ and $v_i$.
    \item The voter leaves the booth with their receipts.
    \item The DRE merges both parts of the receipt in a single part $(i:(U_i,V_i,E_i,W_i,P_{WF}\{$  $E_i\},P_K\{s_1\}),(\text{confirmed},P_K\{s\})$ and creates a block to mine it at its blockchain. The transaction is sent to BB.
    \item The voter verifies that their receipts match those on the BB.
    \end{enumerate}.
\item[] \textit{Verification by blockchain:}
This phase involves the DRE., BB and the underlying blockchain.\\
The blockchain consensus verifies all the zero knowledge proofs in a receipt before adding it to the blockchain. It also verifies consistency of $r_i$ and $v_i$ in case of an audited ballot. The blockchain drops a receipt if the verification of any of its zero knowledge proofs fails.
\item[] \textit{Tallying Phase:}
This phase involves the DRE machine, BB, the blockchain and the public.
\begin{enumerate}
    \item[i.] The DRE posts on the BB the final tally, $t$, final sum $s$ and $m$.
    \item[ii.] The public:
    \begin{enumerate}
        \item[1)] verify all the well-formedness proofs on the BB (\textbf{well-formedness verification})
        \item[2)] verify that for all the audited ballots on the BB: the first part of the receipt,\\ $(U_i, V_i, E_i, W_i, P_{WF}\{E_i\}, P_K\{s_1\})$, is consistent with $r_i$ and $v_i$.
        \item[3)] verify that all the following equations hold (tally verification)
        $$\Pi_{j\in\mathbb{C}}U_j = g_1^s, \Pi_{j\in\mathbb{C}}V_j = g_2^s, \Pi){j\in\mathbb{C}}E_j = h^s g_1^t, \Pi_{j\in\mathbb{C}}W_j = c^s d^m$$
    \end{enumerate}
\end{enumerate}
\end{itemize}
\subsection{Multiple-Candidate Case}
For the case when there are $n \geq 3$ candidates, we consider an upper bound, say $N$, on the number of voters and will encode the vote for the $j$-th candidate as $v-I = N^{j-1}$. We will therefore have to extend the $i$-th ballot to be of the form
$$((U_i, V_i, E_i, W_i, P_{WF}\{E_i\}, P_K\{s_1\}), (\text{audited}, r_i, v_i))$$ in case of audit, or
$$((U_i, V_i, E_i, W_i, P_{WF}\{E_i\}, P_K\{s_1\}), (confirmed, P_K\{s\}))$$ in case of confirmed vote, where $E_i = h^{r_i}g_1^{N^{j-1}}$. The well-formedness proof $P_{WF}\{E_i\}$ will be 1-out-of-n disjunctive proof and can be stated as:
$$P_{WF}\{E_i: g_1, g_2, c, d, h, U_i, V_i, W_i\} = P_K\{r_i: \lor_{j=1}^n((U_i = g_1^{r_i})\land(V_i = g_2^{r_i})\land(E_i/g_1^{N^{j-1}} = h^{r_i})\land(W_i =$$
$(cd^{\alpha_i})^{r_i}))\}$

\section{Conclusion}
This report presents a method of voter verification which, although not entirely new, improves upon previously established methods and tries to make them more scalable and cost-effective, especially with a view to the situation in countries such as India. The entire protocol can be implemented with the help of a mobile app on the client(voter) side, and a well-structured database on the government's side. Hence the only real prerequisite we are assuming is the availability of smartphones, which in today's age is fast becoming a reality. Our proposed method aims to be more or less easily implementable on a large scale, while still maintaining its trustless and error-free nature.
\section{References}
[1] Panja, Somnath; Roy, Bimal K.; \textit{A secure end-to-end verifiable e-voting system using zero knowledge based blockchain}

[2] Grassi, Lorenzo; Khovratovich, Dmitry, Rechberger, Christian; Roy, Arnab; Schofnegger, Markus; \textit{POSEIDON: A New Hash Function for Zero-Knowledge Proof Systems}

[3] \textit{Iden3 Documentation}, \textbf{Link: }https://docs.iden3.io

[4] Bottinelli, Paul; \textit{Security Considerations of zk-SNARK Parameter Multi-Party Computation}, \textbf{Link: }https://research.nccgroup.com/2020/06/24/security-considerations-of-zk-snark-parameter-multi-party-computation

[5] \textit{Semaphore Documentation}, \textbf{Link: }https://semaphore.appliedzkp.org/docs/technical-reference

[6] Fiore, Dario; \textit{Zero-Knowledge Proofs for Set Membership}, \textit{IN THE ART OF ZERO KNOWLEDGE}, Feb 27, 2020

[7] Petkus, Maksym; \textit{Why and How zk-SNARK works}
\end{document}